\documentclass{article}
\usepackage{spconf,amsmath,graphicx,tabularx,color,blindtext,booktabs}
\usepackage{hyperref}

\DeclareMathOperator{\sinc}{sinc}

\title{ACOUSTIC MODEL ADAPTATION FROM RAW WAVEFORMS WITH SINCNET}
\name{Joachim Fainberg, Ond\v{r}ej Klejch, Erfan Loweimi, Peter Bell, Steve Renals\thanks{This work was partially supported by a PhD studentship funded by Bloomberg, by the EU H2020 project ELG (grant agreement 825627), and by the EPSRC project SpeechWave (EP/R012180/1).}}
\address{Centre for Speech Technology Research, University of Edinburgh, United Kingdom}

\begin{document}
\maketitle
\begin{abstract}
Raw waveform acoustic modelling has recently gained interest due to neural networks' ability to learn feature extraction, and the potential for finding better representations for a given scenario than hand-crafted features. SincNet has been proposed to reduce the number of parameters required in raw-waveform modelling, by restricting the filter functions, rather than having to learn every tap of each filter. We study the adaptation of the SincNet filter parameters from adults' to children's speech, and show that the parameterisation of the SincNet layer is well suited for adaptation in practice: we can efficiently adapt with a very small number of parameters, producing error rates comparable to techniques using orders of magnitude more parameters.
% - raw waveform argument: hand-crafted feature extraction not necessarily optimal for a given task
% - adaptation argument: filterbanks vary by task
% - num parameters argument: easy to overfit, raw waveform models which learn every tap of the filters large amounts of parameters, yet this might be where we want to adapt if we adapt for low level acoustic features such as speaker acoustics
% - SincNet: enables the training of acoustic models from raw waveforms in a parameter efficient manner, interpretable filters
% - Experiment and explore with adaptation
% - VTLN: compensate for vocal tract length between speakers

% SincNet enables the training of acoustic models from raw waveforms in a parameter efficient manner. We study the adaptation of the filter parameters in the SincConv layers to speakers and noise factors. We show that we can efficiently adapt to small amounts of data in a robust manner. We study the shifts of the filters, and relate them to e.g. VTLN.
\end{abstract}
\section{Introduction}
Automatic speech recognition models have in recent years obtained impressive word error rates (WERs)~\cite{xiong2018microsoft}. A key component to improving performance is to reduce mismatch between the acoustic model and test data, by explicit adaptation or normalisation of acoustic factors (e.g. \cite{swietojanski2016learning,gales1998maximum,klejch2018learning,sim2018domain,fainberg2017factorised}). Methods such as Vocal Tract Length Normalisation (VTLN)~\cite{lee1996speaker}, which aims to mitigate large variations in individual speakers acoustics, scales the filterbank in standard feature extraction. There has, however, been a growing interest in reducing the amount of hand-crafted feature extraction that is required for acoustic modelling of speech~\cite{tuske2014acoustic,sainath2015learning,ravanelli2018speaker,takeda2018multi}. The motivations to learn part, or all, of the feature extractor range from aiding interpretability~\cite{tuske2014acoustic,ravanelli2018speaker}, to obtaining more optimal representations for the task at hand~\cite{sainath2013learning}. Jaitly and Hinton~\cite{jaitly2011learning}, for example, argued that low-dimensional, hand-crafted features, such as Mel-frequency cepstral coefficents (MFCCs), may lose relevant information that is otherwise present in the original signals.

From raw time-domain waveforms, convolutional neural networks (CNN) have shown promising results
~\cite{tuske2014acoustic,sainath2015learning,palaz2015analysis,hoshen2015speech}. It has even been demonstrated that it is possible to learn band-pass beamformers from multi-channel raw waveforms~\cite{hoshen2015speech}, and a feature extractor learned from raw frequency representations of speech has been shown to outperform conventional methods~\cite{ghahremani2018acoustic}. Their interpretability, however, is sometimes limited, and it is not always clear how to apply existing adaptation techniques. In a recent approach called SincNet~\cite{ravanelli2018speaker}, Ravanelli and Bengio propose to constrain the CNN filters learned from raw time-domain signals, by requiring each kernel to model a rectangular band-pass filter. The authors show that this yields improved efficiency, and that the filters are more easily interpretable.

In this paper we propose to make use of these characteristics for the adaptation of raw waveform acoustic models: we would like efficient, compact representations that are quick to estimate and cheap to store. We explore whether we can obtain this by adapting the cut-off frequencies, and the gains of the filters in SincNet.  This layer may be particularly well suited for speaker adaptation, as the lower layers are known to carry more speaker information than the other layers~\cite{mohamed2012understanding,swietojanski2014learning}. We will show in Section~\ref{sec:methods_rel} that adapting this parameterisation of the CNN filters has similarities with, and crucial differences from, VTLN, feature-space Maximum Likelihood Linear Regression (fMLLR)~\cite{gales1998maximum}, and Learning Hidden Unit Contributions (LHUC)~\cite{swietojanski2016learning}. VTLN has been used to mitigate large variations in vocal tract length for the recognition of children's speech~\cite{potamianos2003robust}. In our experiments we adapt from adults' to children's speech and show that we obtain VTLN-like scaling functions of the filter frequencies. 

There are related approaches in literature that aim to learn, and update filterbanks on top of e.g. raw spectra~\cite{sainath2013learning,sainath2015learning,seki2017deep,seki2018rapid}. As argued in these papers, fixed filterbanks may not be an optimal choice for a particular task. Sailor and Patil~\cite{sailor2016novel} indeed showed that their proposed convolutional restricted Boltzmann machine (RBM) model learns different centre frequencies depending on the task at hand. Our work is perhaps most closely related to Seki et al.~\cite{seki2018rapid}, who proposed to adapt a filterbank composed of differentiable functions such as Gaussian or Gammatone filters. They demonstrated more than 7\% relative reductions in WER when adapting to speakers in a spontaneous Japanese speech transcription task. Our work differs in that we propose to adapt the SincNet layer, which operates on raw waveforms, rather than power spectra.

We review SincNet and related methods in Section~\ref{sec:methods}. Section~\ref{sec:exp_setup} presents the experimental setup, with results in Section~\ref{sec:results}. Section~\ref{sec:conclusion} concludes the paper.

%\section{Methods}
\section{SincNet}
\label{sec:methods}
The idea of SincNet~\cite{ravanelli2018speaker} is to use rectangular band-pass filters in place of standard CNN filters for raw-waveform acoustic models. A rectangular filter with lower and upper cut-off frequencies $f_l$ and $f_u$ has the following time-domain representation, represented as the difference between two low-pass filters:
\begin{equation}
    g[n, f_u, f_l] = 2f_u \sinc (2\pi f_u n) - 2f_l \sinc (2\pi f_l n).
\end{equation}

\begin{figure}
    \centering
    \includegraphics[width=0.8\columnwidth,trim=0 6 0 0,clip]{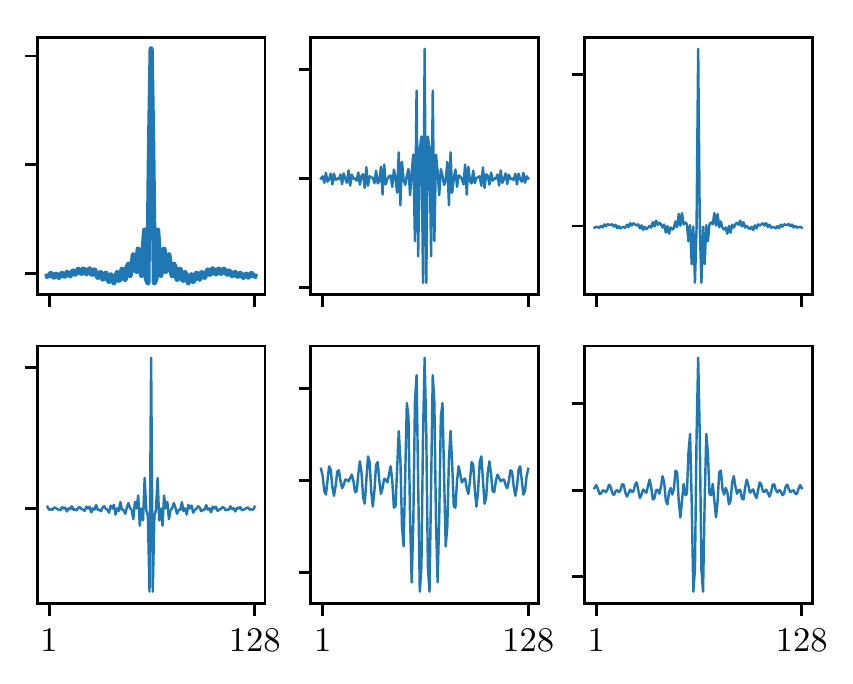}
    \caption{Examples of learned bandpass filters in the time-domain.}
    \label{fig:sinc_filters_example}
\end{figure}

Consequently, the number of parameters per filter is reduced from having to model every tap of each filter (i.e. the filter length) to only having to model two: the cut-off frequencies of the filters, regardless of filter length. An example of learned filters are shown in Figure~\ref{fig:sinc_filters_example}. As in~\cite{ravanelli2018speaker} we use Hamming windows~\cite{oppenheim1975digital} to smooth discontinuities towards the edges:
\begin{equation}
    w[n] = 0.54 - 0.46 \cos(\frac{2\pi n}{L}),
\end{equation}
where $L$ is the filter length. Consequently, the final forward pass for speech input $x[n]$ with one filter is:
\begin{equation}
    y[n] = x[n] * g_w[n, f_u, f_l] = x[n] * w[n]\,g[n, f_u, f_l].
\end{equation}

A set of filters becomes a learnable filterbank of approximately rectangular filters. A related method by Seki et al.~\cite{seki2017deep} replaced the standard Mel-filterbank during feature extraction of Mel-frequency cepstral coefficients (MFCCs) with differentiable Gaussian filters on top of power spectra, enabling the learning of centre frequencies, bandwidths and gain. SincNet also learns a filterbank, but in the time-domain on raw waveform features. For SincNet, Ravanelli and Bengio~\cite{ravanelli2018speaker} chose not to explicitly model the gain of each filter, as it can be readily learned by later parts of the neural network. 

\subsection{Relationship with VTLN, fMLLR and LHUC}\label{sec:methods_rel}
\begin{figure}
    \centering
    \includegraphics[width=0.95\columnwidth,trim=2 6 0 2,clip]{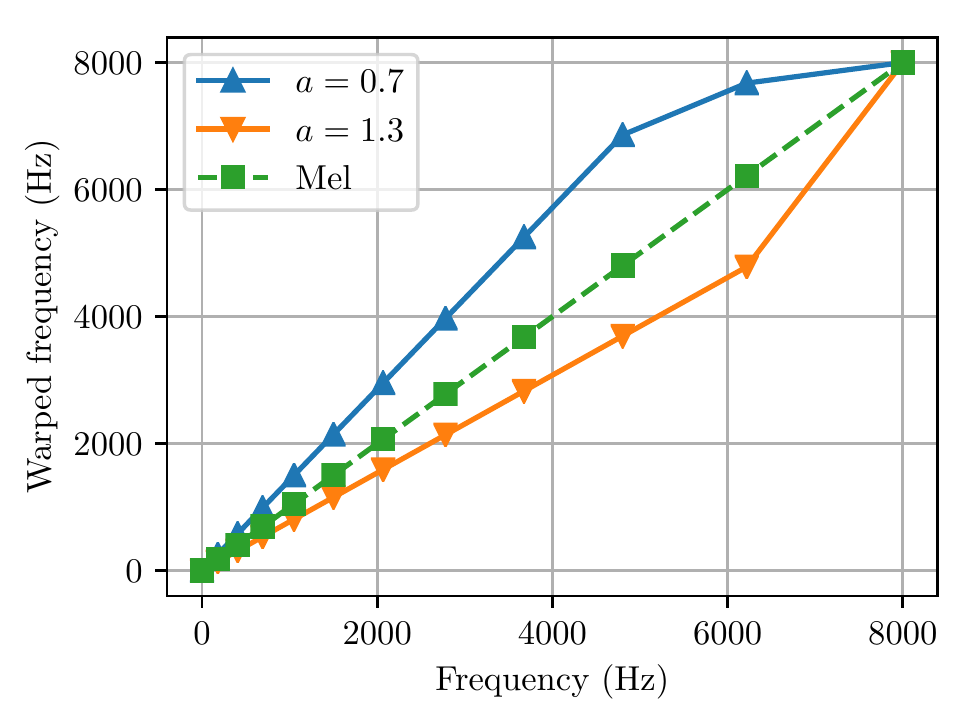}
    \caption{Example of piece-wise linear scaling functions for VTLN. $\alpha=1$ would equal the Mel line.}
    \label{fig:vtln_example}
\end{figure}

A learnable filterbank has close relationships with other well-known methods, as also previously highlighted by Seki et al.~\cite{seki2017deep}. In this paper we suggest to update the SincNet filterbank
%on a speaker-level. 
for each speaker.
This strongly resembles VTLN~\cite{lee1996speaker}, which aims to compensate for varying vocal tract lengths among speakers. It accomplishes this by scaling, or warping, the centre frequencies of the filters in the Mel-filterbank. Consequently, adapting the parameters of the SincNet layer resembles VTLN with a few key differences:

\begin{enumerate}
\item SincNet operates in the time-domain, and uses corresponding rectangular filters rather than triangular filters as in the Mel-filterbank;
\item VTLN typically uses a scaling function that is assumed to be piece-wise linear with a single slope parameter, $\alpha$ (as shown in Figure~\ref{fig:vtln_example}), whilst if adapting SincNet, the effective learned scaling functions are
less constrained.
%not constrained to be piece-wise linear;
\item The slope parameter $\alpha$ is typically determined with a grid search (although, there exist more sophisticated methods such as gradient search~\cite{panchapagesan2006multi}). With SincNet we can learn the scaling function using gradient descent.
\end{enumerate}

In the original SincNet formulation~\cite{ravanelli2018speaker} the gains of the filters are held fixed. Downstream layers can learn to scale the contributions of the filters. However, the filter gains may be suitable targets for adaptation for which we would like to attribute importance to the output of individual filters with a small number of parameters. This has similarly been done with learnable filterbanks in traditional feature extraction pipelines~\cite{seki2018rapid}. We also briefly note that if we were to scale the gain of each filter, then this would correspond to a version of feature-space Maximum Likelihood Linear Regression (fMLLR)~\cite{gales1998maximum} with a diagonal matrix and no bias, or similarly to Learning Hidden Unit Contritutions (LHUC)~\cite{swietojanski2016learning} which scales the output of each neuron by a scalar $r^{(i)}$ for filter $i$:
\begin{equation}
    y^{(i)}[n] = r^{(i)} \sum_{l=0}^{L-1} x[l]\,g_{w}^{(i)}[n-l].
\end{equation}
Clearly, we can view the vector of scalars, $\mathbf{r}$, as either scaling the features, the gain of the filters, or the output of the layer.

\section{Experimental setup}\label{sec:exp_setup}
Our baseline models are built using the AMI Meeting Corpus~\cite{carletta2007unleashing}, which contains about 70 hours of training data from fictitious design team meetings. We use the individual head-mounted
microphone (IHM) stream, and trained HMM-GMM systems using Kaldi~\cite{povey2011kaldi} following the recipe for AMI\footnote{\url{github.com/kaldi-asr/kaldi/tree/master/egs/ami}}.

As adaptation data we use children's speech from the British English PF-STAR corpus~\cite{batliner2005pf_star}, which in total consists of roughly 14 hours of data of read children's speech. The children are aged between 4--14, with the majority being 8--10 years old. The data contains a fair amount of mispronunciation and hesitation, making recognition challenging. It is clearly mismatched from the AMI data, in terms of what is spoken, the speaking style, and the acoustics of the speakers~(see e.g. \cite{fainberg2016improving}).

\subsection{SincNet acoustic model}\label{sec:exp_setup_model}
The neural network acoustic model is detailed in Table~\ref{tab:model}. The first layer consists of 40 Sinc filters, each with length 129 as has been used previously in a speech recognition task~\cite{loweimi2019interpretable}. We experimented with different methods of initialising  the upper and lower frequencies of each filter as follows:
\begin{enumerate}
    \item Mel-scale~\cite{o1987speech} as in Ravanelli and Bengio~\cite{ravanelli2018speaker}: the lower frequencies of the filters are linearly interpolated between the corresponding mels of $f_{min}$ and $f_{max}$, with $f_u[i] = f_l[i-1]$;
    \item Uniformly at random in the same range. This effectively becomes linear when sorted by centre frequency: $f_l \sim \mathcal{U}(f_{min}, f_{max})$,
    and $f_u[i] = f_l[i-1]$;
    \item  Flat with $f_l=f_{min}$ and $f_u=f_{min} + b$ for each filter (randomness is induced by the layers above).
\end{enumerate}
For each scheme we set $f_{min}=30\,\text{Hz}$ and $f_{max}=sr/2 - (f_{min}+b)$, where $sr$ is the sampling rate (16\,kHz in our experiments), and $b=50\,\text{Hz}$ is the minimum bandwidth.
The remaining layers consist of six 1-D convolution layers with ReLUs, each with 800 units. Kernel sizes and dilation rates are shown in Table~\ref{tab:model}. Batchnorm (BN) layers are interspersed throughout. The final softmax layer outputs to 3,976 tied states.

The models are trained for 6 epochs using Adam~\cite{kingma2014adam} with a batch size of 256 and a learning rate of 0.0015, unless noted otherwise. The waveforms are sampled as in~\cite{ravanelli2018speaker,loweimi2019interpretable}: we use 200\,ms windows with a shift of 10\,ms, i.e. the input size to the network is $16000 * 0.200 = 3200$. We implemented and trained the models using Keras~\cite{chollet2015keras} and Tensorflow~\cite{abadi2016tensorflow}. We decode and score with Kaldi~\cite{povey2011kaldi}. Our experimental code is publically available\footnote{\url{github.com/jfainberg/sincnet_adapt}}.

\begin{table}[th]
  \centering
  \begin{tabular}{c l c c c l}
    \toprule
    \multicolumn{1}{l}{\textbf{\#}} & 
      \multicolumn{1}{l}{\textbf{Type}} & \multicolumn{1}{c}{\textbf{Dim}} &
      \multicolumn{1}{c}{\textbf{Size}} & \multicolumn{1}{c}{\textbf{Dil}} &
      \multicolumn{1}{l}{\textbf{Params}}\\
    \midrule
    1 & SincConv & 40 & 129 & - & 80\\
    - & MaxPooling & - & 3 & - \\
    2 & BN(ReLU(Conv)) & 800 & 2 & 1 & 68,000\\
    - & MaxPooling & - & 3 & - \\
    3 & BN(ReLU(Conv)) & 800 & 2 & 3 & 1,284,000\\
    - & MaxPooling & - & 3 & - \\
    4 & BN(ReLU(Conv)) & 800 & 2 & 6 & 1,284,000\\
    - & MaxPooling & - & 3 & - \\
    5 & BN(ReLU(Conv)) & 800 & 2 & 9 & 1,284,000\\
    - & MaxPooling & - & 2 & - \\
    6 & BN(ReLU(Conv)) & 800 & 2 & 6 & 1,284,000 \\
    7 & ReLU(Conv) & 800 & 1 & 1 & 640,800 \\
    8 & Softmax(Conv) & 3976 & 1 & 1  & 3,184,776 \\
    \bottomrule
  \end{tabular}
  \caption{Model topology. In total there are 9,029,656 parameters (including batchnorm).}\label{tab:model}
\end{table}

\subsection{Language model}
As the acoustic and language models for AMI are greatly mismatched to PF-STAR, we  interpolate the standard AMI language model based on AMI and Fisher~\cite{cieri2004fisher} data, with the training data from PF-STAR. This is similar to other literature working with PF-STAR~\cite{dubagunta2019improving}. We note, however, that there is some overlap in the sentences between training and test sets for PF-STAR, i.e. training a LM on the training set causes some data leakage.  For this paper we believe this is acceptable, given that we are interested in the acoustic model mismatch. Without the biased LM, the combined effect of a mismatched LM, and a mismatched AM, produced WERs greater than 90\% in our preliminary experiments.

We estimate a 3-gram LM with Kneser-Ney discounting on the PF-STAR training set using the SRILM toolkit~\cite{stolcke2002srilm}. This is interpolated with the AMI model, giving the latter a weight of $0.7$. The vocabulary is restricted to the top 150k words from an interpolated 1-gram model. Finally, we prune the interpolated model with a threshold of $10^{-7}$.

\section{Results}\label{sec:results}
The results with our models trained on AMI are shown in Table~\ref{tab:baseline_filters}. The various initialisation schemes produce quite similar WERs, but, perhaps surprisingly, the Mel-initialisation performs least well. The differences are, however, less than 2\% relative. Overall these numbers are roughly 5 percentage points worse than those produced with cross-entropy systems in the corresponding Kaldi recipe for AMI. A key difference may be the use of speed perturbation for data augmentation~\cite{ko2015audio}. Our models are slow to train, but we propose improving training speed as an area for future work.

\begin{figure}[th!]
    \centering
    \includegraphics[width=0.98\columnwidth,trim=2 6 0 2,clip]{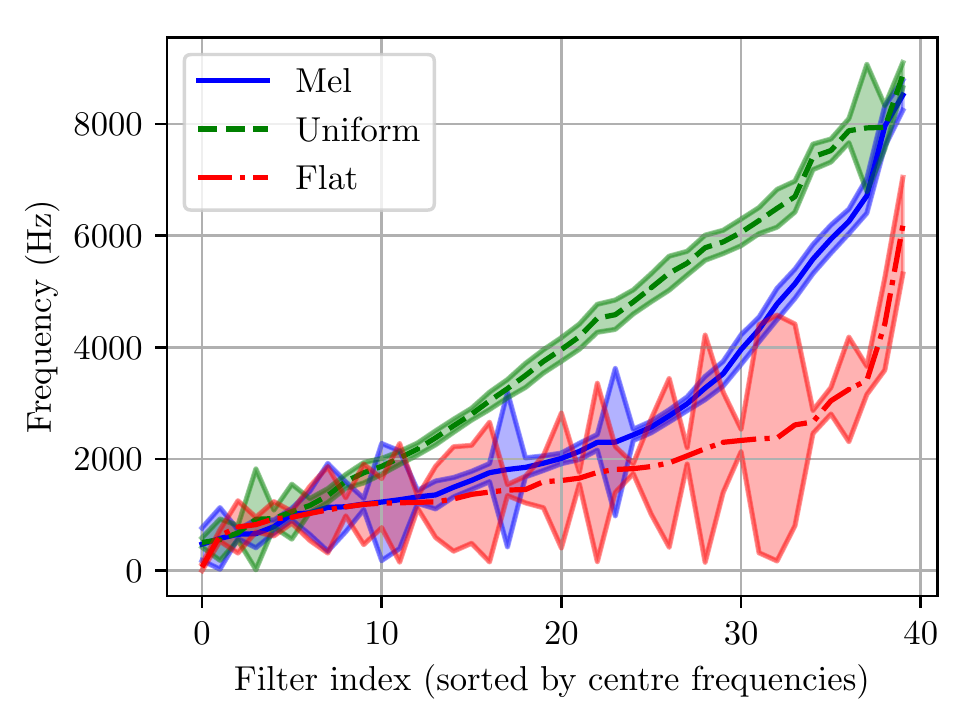}
    \caption{Upper and lower learned frequencies per filter with different initialisation schemes, after six epochs of training on AMI. In contrast to Mel and Uniform, Flat is forced to change in order to extract different information in each filter.}
    \label{fig:filter_response}
\end{figure}

\begin{figure}[th!]
    \centering
    \includegraphics[width=0.98\columnwidth,trim=2 6 0 2,clip]{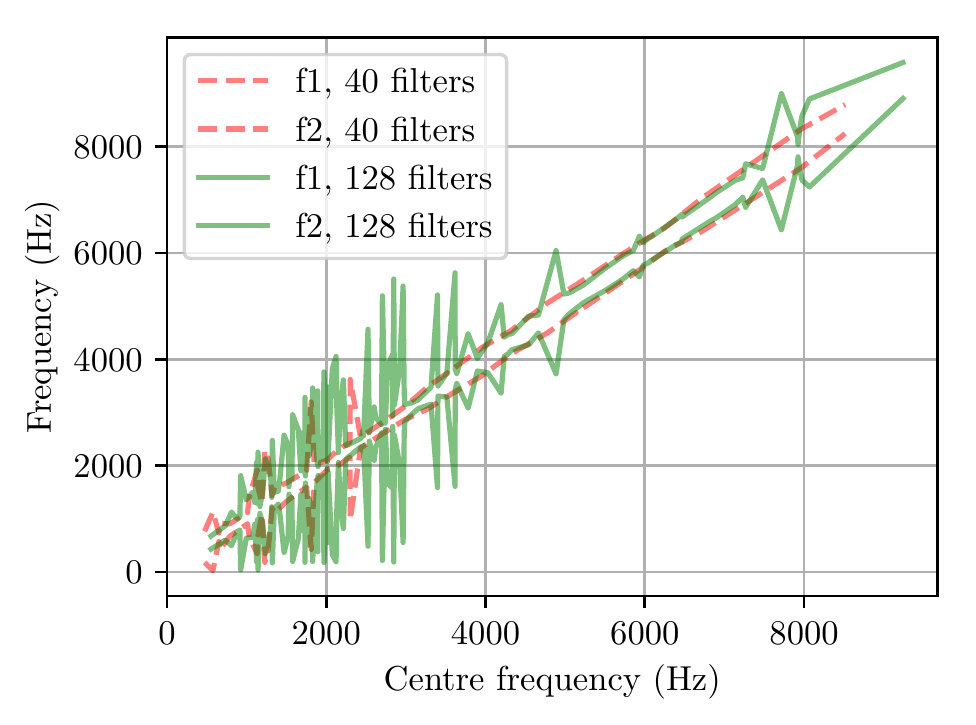}
    \caption{Cut-off frequencies per filter for filterbanks with 40-filters or 128-filters after training with a Mel-initialised filterbank, plotted against their respective centre frequencies for comparison. We observed similar WERs, with simply more noisy bandwidths, with an increased number of filters.}
    \label{fig:filter_response_40v128}
\end{figure}

Figure~\ref{fig:filter_response} demonstrates how the initialisation schemes lead to different final responses in the filters. The flat initialisation is essentially forced to change significantly, otherwise each filter would extract identical information. After training it begins to approximate a Mel-like curve. This is in line with similar research~\cite{sainath2015learning,sailor2016novel}. We noted in Section~\ref{sec:exp_setup_model} that the uniform initialisation effectively creates a linear initialisation. It remains largely linear after training, with some shifts in higher frequencies, and changes to bandwidths. Note that each response is markedly different, yet the corresponding WERs are similar. This may be explained by
%, as noted by Seki et al., 
the ability of the downstream network to learn to use the extracted features in different ways~\cite{seki2018rapid}. We also experimented with a larger number of filters than 40 (e.g. 128), but saw no benefit to WERs; instead, the filter bandwidths become quite erratic, as demonstrated in Figure~\ref{fig:filter_response_40v128}.

% \todo{discuss distribution of subband filters in LF vs HF regions, similar to Mel scale? \cite{sailor2016novel}}

We use the model trained from a flat initialisation 
for our further experiments.
%from now on.

\begin{table}[th]
  \centering
  \begin{tabular}{l c c}
    \toprule
    \multicolumn{1}{l}{\textbf{Initialisation}} & 
      \multicolumn{1}{c}{\textbf{Eval}} & \multicolumn{1}{c}{\textbf{Dev}}\\
    \midrule
    % 40       & Mel   & $31.6$   &  3 \\ % exp/tdnn_am_ondrej_small_40
    % 128      & Mel   & $31.8$   &  3 \\ % exp/tdnn_am_ondrej_small
    % 40       & Flat   & $31.4$   &  4 \\ % exp/ami_sinc_40_uniinit
    % 128      & Flat   & $31.5$   &  4 \\ % exp/ami_sincnet_flatinit
    % 40       & Uni   & $30.9$   &  4 \\ % exp/ami_sinc_40_flatinit
    % 128      & Uni   & $31.0$   &  4 \\ % exp/ami_sinc_128_uniinit
    Mel   & $30.6$ & $28.0$  \\ % exp/ami_sinc_40_mel_6epochs
    Flat  & $30.2$ & $28.0$  \\ % exp/ami_sinc_40_flat_6epochs
    Uni   & $30.3$ & $27.9$  \\ % exp/ami_sinc_40_uni_6epochs
    \bottomrule
  \end{tabular}
  \caption{Results (\% WER) on AMI with various filter initialisation.}
  \label{tab:baseline_filters}
\end{table}

\subsection{Domain adaptation to children's speech}
We next investigate supervised domain adaptation of the SincConv layer from AMI to PF-STAR (from adults' to children's speech). As shown in Table~\ref{tab:childrens_speech}, the AMI model is initially highly mismatched with PF-STAR, with a WER of $68.19\%$, which aligns with what is expected from the literature~\cite{potamianos2003robust}. For reference we include a model trained from scratch on PF-STAR, which obtains $20.46\%$ WER. Adapting the SincConv layer of the AMI model for a single epoch to the training set of PF-STAR reduces the error rate to $31.65\%$. We include numbers showing the effect of updating the statistics of the batchnorm layers. Freezing the batchnorm layers demonstrates that the primary improvement comes from adapting the 80 parameters in the SincConv layer. We freeze the batchnorm layers in all experiments that follow. This experiment shows that we can effectively adapt a very small number of parameters in the model, improving the out-of-domain model by over $50\%$ relative, and coming within 12 percentage points of a model trained from scratch with all 9M parameters. Adapting the SincConv layer amounts to adapting less than 0.0009\% of the total number of parameters in the model (see Table~\ref{tab:model}).

\begin{table}[th]
  \centering
  \begin{tabular}{ l c c c}
    \toprule
    \multicolumn{1}{l}{\textbf{Model}} & 
      \multicolumn{1}{c}{\textbf{WER (\%)}}\\
    \midrule
    % Unadapted 40-Uni* (not matching model) & $66$ \\ % exp/ami_sinc_40_uniinit/decode_pfstar_test -> other model decoding
    AMI              & $68.19$ \\ % exp/ami_sinc_40_flat_6epochs//decode_pfstar_test
    PF-STAR          & $20.46$ \\ % exp/pfstar_sinc_40_mel_20epochs
    \midrule
    Adapt+batchnorm  & $29.90$ \\ % exp/ami_sinc_40_flat_6epochs/adapt_pfstar/decode_pfstar_test ep train / 2
    Adapt--batchnorm & $31.65$ \\ % exp/ami_sinc_40_flat_6epochs/adapt_pfstar_nobn/decode_pfstar_test ep train / 2
    \bottomrule
  \end{tabular}
  \caption{Results (\% WER) on the PF-STAR test set, after adapting the AMI model for one epoch. A PF-STAR model is shown for reference. The models have 40 filters that were originally initialised flat.}\label{tab:childrens_speech}
\end{table}

\begin{figure*}[t]
    \centering
    \includegraphics[width=1.00\textwidth,trim=2 6 0 2,clip]{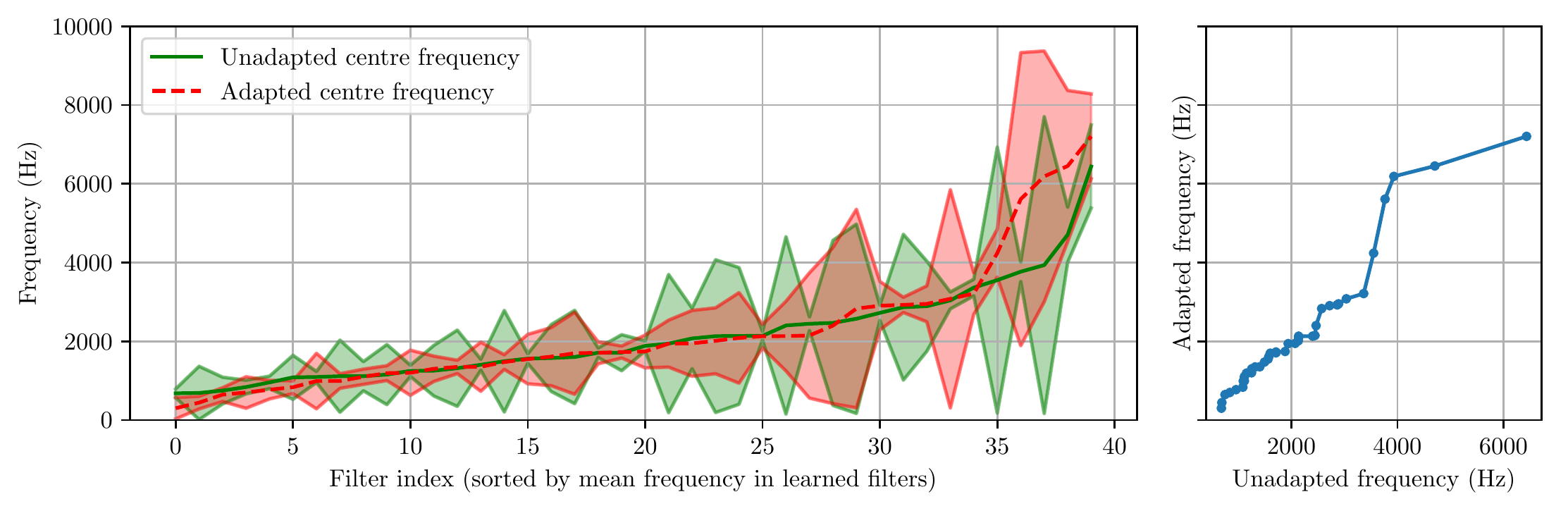}
    \caption{(Left) SincNet filters before and after domain adaptation to PF-STAR. (Right) SincNet centre frequency adapted vs unadapted (e.g. VTLN-like function).}
    \label{fig:domain_vtln}
\end{figure*}

\begin{figure}
    \centering
    \includegraphics[width=0.95\columnwidth,trim=2 6 0 2,clip]{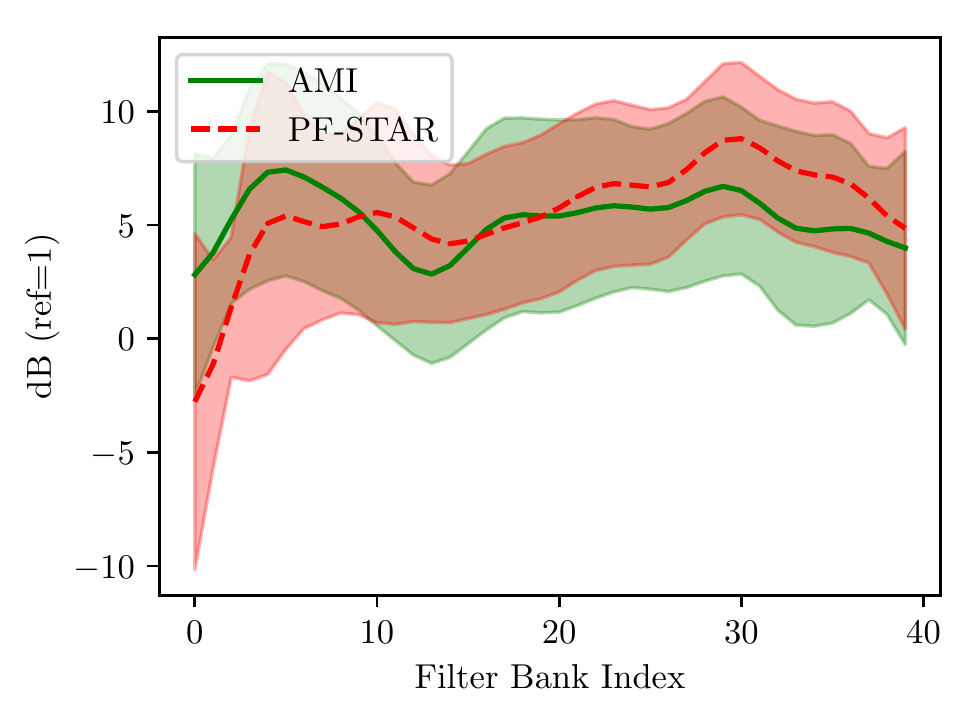}
    \caption{Average log-mel filterbank spectra for random subsets of AMI and PF-STAR data. The shaded region denotes mean plus-minus one standard deviation.}
    \label{fig:spectrogram_comparison}
\end{figure}

Figure~\ref{fig:domain_vtln} shows that adapting the SincConv layer shifts the upper frequency distribution of the filters, and their bandwidths. This is reflected in the corresponding VTLN function. This suggests that the model has adjusted to higher frequency content in the children's speech data. Figure~\ref{fig:spectrogram_comparison} shows average power spectra from a the corresponding log-mel filterbank features of AMI and PF-STAR which supports this notion.

We note that the VTLN-like function is nearly piecewise-linear; i.e. similar to the assumptions made during typical use of VTLN. However, it was here obtained through backpropagation instead of grid-search or other methods.

Table~\ref{tab:sup_domain_by_utt} demonstrates the effect of the number of adaptation utterances. As the amount of data increases adapting all parameters (excluding SincConv) produces lower error rates, as should be expected. The models begin to diverge at about three utterances (roughly 1 minute for PF-STAR).

\begin{table}[th]
  \centering
  \begin{tabular}{ l c c c c c }
    \toprule
    \multicolumn{1}{l}{\textbf{Adapt/utts}} & 
      \multicolumn{1}{c}{\textbf{0}} &
      \multicolumn{1}{c}{\textbf{1}} & 
      \multicolumn{1}{c}{\textbf{2}}& 
      \multicolumn{1}{c}{\textbf{3}}&
      \multicolumn{1}{c}{\textbf{20}}\\
    \midrule
    Sinc      & $68.19$ & $56.67$ & $47.87$ & $40.36$ & $31.06$ \\
    All--Sinc & $68.19$ & $56.70$ & $47.42$ & $35.89$ & $21.34$ \\ 
    \bottomrule
  \end{tabular}
  \caption{Results (\% WER) on the PF-STAR test set given number of adaptation utterances when adapting from the AMI model. As the amount of data increases, adapting all parameters surpass only adapting the SincConv layer.}\label{tab:sup_domain_by_utt}
\end{table}

\subsection{Speaker adaptation}
\begin{figure}[th!]
    \centering
    \includegraphics[width=1.00\columnwidth,trim=2 6 0 2,clip]{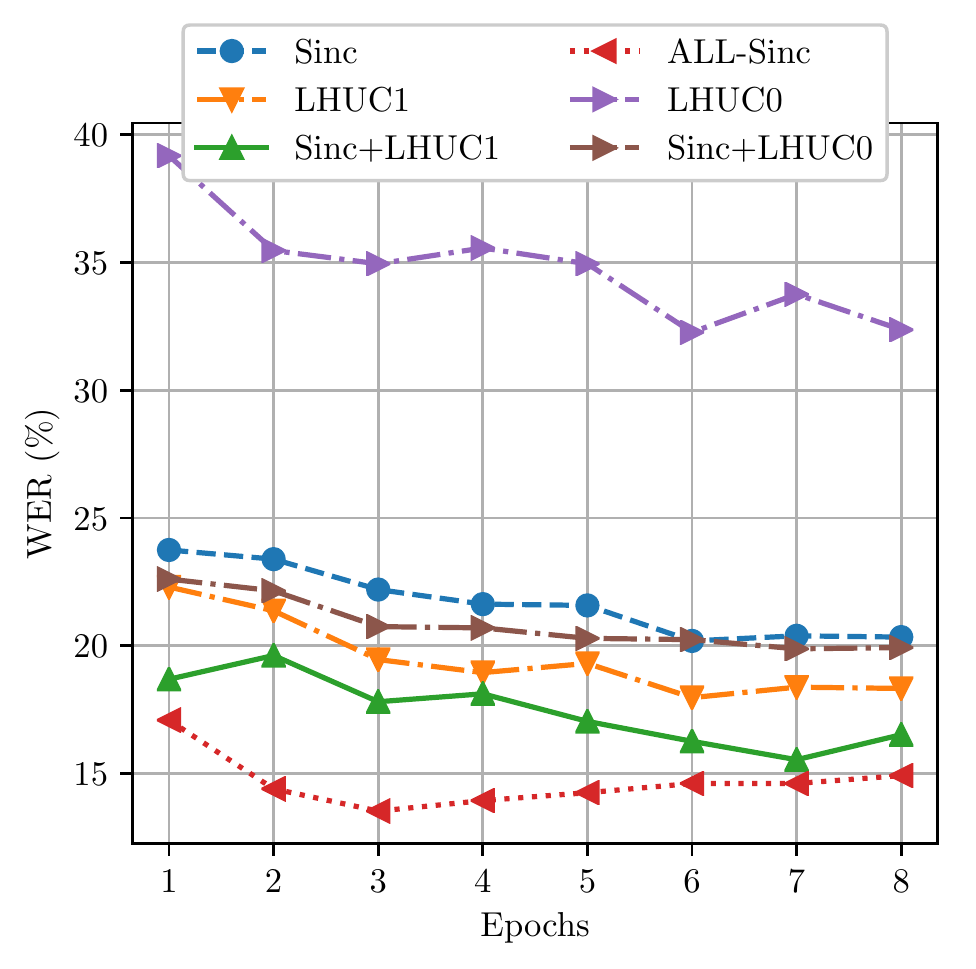}
    \caption{Speaker adaptation over epochs with various techniques. The unadapted model obtains 59.06\% WER.}
    \label{fig:comparison_spk_adapt}
\end{figure}

A more realistic, practical scenario, is to adapt to a few utterances obtained per speaker. In these experiments we adapt from the AMI model to 12 individual speakers in PF-STAR's eval/adapt set, testing on the corresponding speakers in eval/test. The results are shown in Figure~\ref{fig:comparison_spk_adapt} which shows the evolution with the number of epochs of adaptation. \texttt{LHUC0} indicates using LHUC on the output of the SincConv layer (40 parameters), and \texttt{LHUC1} is LHUC on the output of the first CNN layer (800 parameters). We use a learning rate of $0.8$ for \texttt{LHUC0} and \texttt{LHUC1}, as LHUC can characteristically use very large learning rates without overfitting~\cite{swietojanski2016learning,klejch2018learning}. When using \texttt{LHUC1} in combination with SincConv, we use the standard $0.0015$ learning rate for SincConv, but a multiplier of $500$ for LHUC. For \texttt{Sinc+LHUC0} we did not find this beneficial and used the same learning rate for both.

The un-adapted WER is 59.06\%. Adapting the 80 parameters of the SincConv layer yields only slightly worse results than \texttt{LHUC1} with 10 times fewer parameters. Interestingly, the two are complementary, as demonstrated by \texttt{Sinc+LHUC1}, and at best produces WERs similar to adapting all 9M parameters (excluding SincConv). \texttt{ALL-Sinc} is more sensitive to overfitting as evident from the figure.

Adapting the gain of the filters improves substantially over the unadapted model, but does not provide similar performance to any of the other approaches. One factor may be that the these parameters were fixed during the training of the baseline model as in~\cite{ravanelli2018speaker}, hence the rest of the network may have compensated by other means. It is, however, complementary with adapting the filterbank frequencies, with \texttt{Sinc+LHUC0} slightly outperforming \texttt{Sinc}. A summary of the results after adapting for eight epochs is shown in Table~\ref{tab:sup_speaker_summary}.

\begin{figure}[th]
    \centering
    \includegraphics[width=1.00\columnwidth,trim=2 6 0 2,clip]{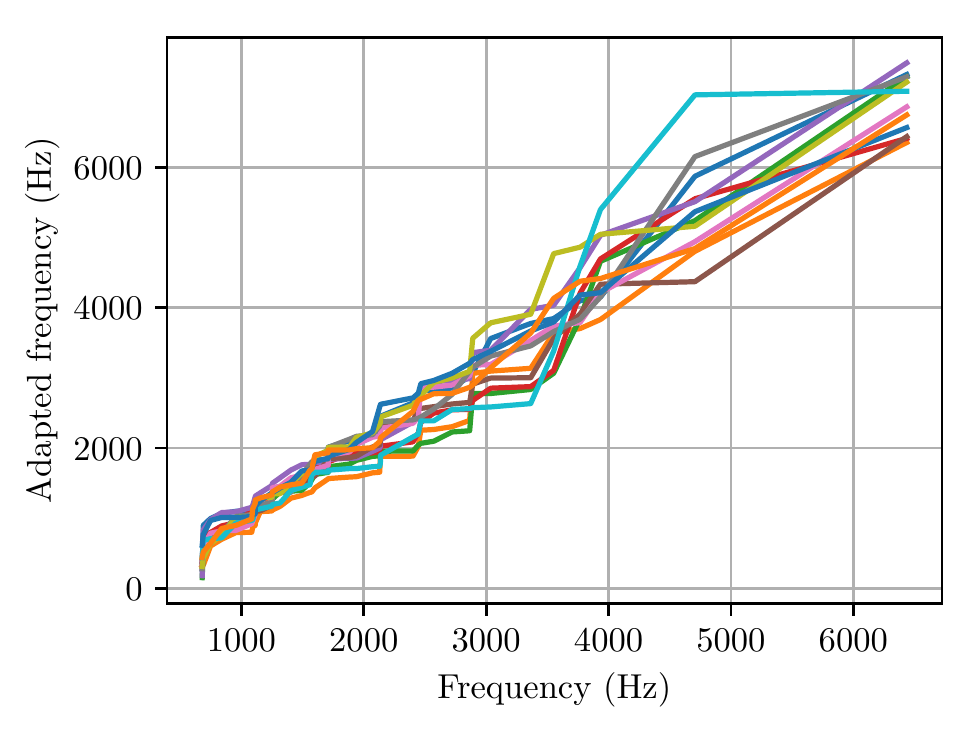}
    \caption{Corresponding VTLN scaling functions for individual speakers (colours) adapted with \texttt{Sinc}. The majority of the scaling occurs in the higher frequencies.}
    \label{fig:sup_spk_adapt_vtln}
\end{figure}

Figure~\ref{fig:sup_spk_adapt_vtln} shows VTLN-like functions obtained from the adapted SincConv layer to each speaker. There is a clear difference between each function, which is in line with what one might expect given the variability of the acoustics of children's data~\cite{potamianos2003robust}.

\begin{table}[th]
  \centering
  \begin{tabular}{ l c c}
    \toprule
    \multicolumn{1}{l}{\textbf{Method}} & 
      \multicolumn{1}{c}{\textbf{WER (\%)}} &
      \multicolumn{1}{c}{\textbf{Params}}\\
    \midrule
    Unadapted            & $59.06$  & -    \\
    \midrule
    \texttt{Sinc}        & $20.34$ & $80$  \\
    \texttt{LHUC0}       & $32.37$ & $40$  \\
    \texttt{Sinc+LHUC0}  & $19.93$ & $120$ \\
    \texttt{LHUC1}       & $18.33$ & $800$ \\
    \texttt{Sinc+LHUC1}  & $16.52$ & $880$ \\
    \texttt{ALL-Sinc}    & $14.92$ & $\sim 9\text{M}$ \\
    \bottomrule
  \end{tabular}
  \caption{Results (\% WER) adapting from AMI to individual speakers in PF-STAR for 8 epochs.}\label{tab:sup_speaker_summary}
\end{table}

% \subsection{Test-time speaker adaptation}
% Adaptation from domain-adapted model, with 1st pass decode. We show that adapting the filterbank frequencies can on average be better than adapting all parameters, which is more susceptible to overfit in this scenario. Two sets of experiments, once from raw AMI model which is very mismatched, and once from domain adapted model.

% \todo{Ideas: learning rate plot; major components / extract VTLN for fewer params; frequency response plot (FFT by filter after Hamming); speaker adaptation AMI}

\section{Conclusions}\label{sec:conclusion}
We have shown that adapting the filterbank frequencies from raw waveforms with SincNet is extremely parameter efficient, obtaining substantial improvements in WERs with a fraction of the total model parameters on a children's speaker adaptation task. It is also complementary with the standard LHUC technique, producing results similar to adapting all 9 million model parameters (excluding the filterbank layer). We also show that the parameterisation of SincNet affords interpretability during adaptation: during domain adaptation to children's speech, the layer learns to pay more attention to higher frequencies. Similarly for speaker adaptation, the change in the filter frequencies effectively resembles VTLN, producing individual scaling functions for each speaker. Finally, we noted that adapting the gain is related to LHUC and fMLLR, and this proved complementary to adapting the filterbank frequencies.

In future work we would like to explore the use of meta-learning (as in \cite{klejch2018learning}) to learn filter-specific learning rates, as well as experimenting with unsupervised, test-time adaptation of the SincConv layer, and exploring the layer's response to noise. Finally, we would like to experiment with the inclusion of  feature-based adaptation methods such as i-vectors~\cite{saon2013speaker}, which has previously been shown to be complementary to model-based adaptation~\cite{samarakoon2016combining}.

\bibliographystyle{IEEEbib}
\bibliography{main}

\end{document}